\documentstyle[multicol,aps]{revtex}
\def\SAS{\Delta_{\rm SAS}}
\begin{document}
\titlepage
\title{
Phase Transition in $\nu=2$ Bilayer Quantum Hall State
}
\author{A. Sawada$^{\rm 1}$, Z.F. Ezawa$^{\rm 1}$, 
H. Ohno$^{\rm 2}$, Y. Horikoshi$^{\rm 3}$, Y. Ohno$^{\rm 2}$,\\
S. Kishimoto$^{\rm 2}$, F. Matsukura$^{\rm 2}$, 
 M. Yasumoto$^{\rm 1}$, A. Urayama$^{\rm 1}$ 
}

\address{$^1$Department of Physics, 
Tohoku University, Sendai 980-77,Japan}

\address{$^2$Research Institute of Electrical Communication, 
Tohoku University, Sendai 980-77, Japan}

\address{$^3$School of Science and Engineering, 
Waseda University, Tokyo 169, Japan}

\maketitle
\vskip5mm
\begin{abstract} 
The Hall-plateau width and the activation energy were measured in 
the bilayer quantum Hall state at filling factor ${\nu}=2$, 1 and 2/3, 
by changing the total electron density and the density ratio 
in the two quantum wells.
Their behavior are remarkably different from one to another.
The ${\nu}=1$ state is found stable over all measured range 
of the density difference,
while the ${\nu}=2/3$ state is stable only around the balanced point.
The $\nu=2$ state, on the other hand, shows a phase transition between 
these two types of the states as the electron density is changed.
\end{abstract}

\begin{multicols}{2}\narrowtext
\setcounter{page}{1}

\setcounter{page}{1}

Electron systems in confined geometries exhibit a rich variety of
physical properties due to the interaction effects in reduced dimensions.
One of the most interesting phenomena is the quantum Hall (QH) effect
in the planar electron system.
In particular, the QH effect in double quantum wells has recently
attracted much attention\cite{PhaseDiagr,SongHe}, 
where the structure introduces additional degrees of freedom 
in the third direction.
Various bilayer QH states are realized\cite{Boebin,DBLYexp}
by controlling system parameters such as 
the strengths of the interlayer and intralayer Coulomb interaction, 
the tunneling interaction and Zeeman effect. 
A good example is the $\nu=1/2$ state\cite{DBLYexp}, 
for which there is no counterpart in the monolayer system.
Here, $\nu$ is the total filling factor. 
It has been also pointed out\cite{EIcoher,WenZee} that 
a novel interlayer quantum coherence (IQC) 
may develop spontaneously in the $\nu=1/m$ state 
with $m$ an odd integer.
Murphy {\it et al.}\cite{Sheena} have reported 
an anomalous activation energy dependence 
in the bilayer $\nu=1$ QH state on the tilted magnetic field,
which is probably one of the signals\cite{EIplasmon,YangMoon} of the IQC.
Another unique feature of this IQC\cite{SSC}
is that the QH state is stable 
at any electron density ratio $n_f/n_b$, where $n_f$ ($n_b$) 
is the electron density 
in the front (back) quantum wells.

So far the QH states at $\nu=$ odd integers have been 
extensively investigated 
from the viewpoints of ``phase transition" 
due to the interlayer correlation.
The $\nu=2$ bilayer QH state has attracted 
less attention 
because it has been thought as a simple "compound" state 
with $\nu=1+1$ made of two non-interacting monolayer $\nu=1$ states.

In this Letter, 
we report the Hall-plateau width and the activation energy 
in three typical bilayer QH states at $\nu = 2/3$, 1, and 2 
by changing the total electron density $n_t=n_f+n_b$ 
as well as the density ratio $n_f/n_b$.
By changing the total density,
we can change the ratio of the interlayer 
to the intralayer Coulomb interactions,
which govern the the basic nature of the bilayer QH states.
Furthermore, the stability of the bilayer QH state, 
which is quite sensitive to the density ratio in general, 
is also tested to clarify the origin of the QH state.
The behaviors of these three states have been found to be 
remarkably different.
The $\nu=2/3$ state is identified as a compound state 
with $\nu=1/3+1/3$, 
whereas the $\nu=1$ state is found to be 
the ``coherent" state\cite{NoteA}.
The $\nu=2$ state, on the other hand, shows a phase transition 
from the compound state to a spin-unpolarized ``coherent" state 
as the interlayer Coulomb interaction is enhanced.

The sample was grown by molecular beam epitaxy on a (100)-ori\-ented 
GaAs substrate, and consists of two modulation doped 
GaAs quantum wells of width 200\,\AA, 
separated by an Al$_{0.3}$\-Ga$_{0.7}$\-As barrier of thickness 
31\,\AA.
Carriers are supplied from the two Si delta-doping sheets 
($5\times10^{11}$\ cm$^{-2}$),
each of which is placed 750\,\AA\ away from one of the quantum wells.
A Hall-bar mesa was formed by conventional photolithography.
Al/Cr Schottky gate electrodes were fabricated on 
both front and back surfaces of the sample so that
the total electron density $n_{\rm t}$ 
and the electron density difference $n_{\rm f}-n_{\rm b}$ 
can be independently controlled 
by adjusting the front $V_{\rm fg}$ 
and the back gate voltage $V_{\rm bg}$.

Measurements were performed with the sample mounted 
in a mixing chamber of a dilution refrigerator. 
The magnetic field with maximum 13.5\,T 
was applied perpendicular to the electron layers. 
Standard low-frequency ac lock-in techniques were used
with currents less than 100\,nA to avoid heating effects.

The electron density in each layer is a key parameter 
in our experiments. 
We obtained the total electron density $n_{\rm t}$ from 
the Hall resistance at low magnetic field,
and the electron densities in each layer 
from Fourier transforms of the Shubnikov-de Haas oscillations.

In Fig.~\ref{Fig1PS}, the Hall resistance is shown 
at various electron density difference $(n_f-n_b)/n_t$
and at a fixed total electron density of 
$1.2\times10^{11}$\,cm$^{-2}$.
Well-developed quantized Hall plateaus are clearly seen 
at $\nu=2/3, 1$ and $2$.
The total electron density of this sample was 
$2.3\times 10^{11}\,$cm$^{-2}$ at zero gate voltage, 
and the mobility was $3.0\times 10^5\,$cm$^2$/Vs 
at temperature $T=30$\,mK. 
The measured tunneling energy gap was 
${\Delta}_{\rm SAS}{\approx}6.8$\,K,
which is in good agreement with 
the value of the self-consistent calculation result 6.7\,K.

We first concentrate on the width of the Hall plateau, 
which is a good indicater of the stability of the QH state.
We have defined it by the width of 
the magnetic field within the $\pm2.5$\,\% range of 
the Hall resistance after subtracting 
the classical Hall resistance\cite{SSC}.
The state is stable when the plateau width is wide, 
and the stability is lost when the width is zero. 
We later correlate the plateau width to the activation energy.

In Fig.~\ref{Fig2PS} we show 
the plateau width of the ${\nu}=2/3$, 1 and 2 QH states  
as a function of the electron density difference 
at various total electron densities.
The data are slightly out of symmetry
with respect to the balanced point.
We expect a perfect symmetry
in an ideal system.

The plateau width of the ${\nu}=2/3$ state has a peak ({\it maximum})
at the balanced point.
As the total electron density {\it decreases},
the plateau width at the balanced point decreases. 
Eventually, the Hall plateau disappears at the total density
{\it less} than $0.8 {\times} 10^{11}$\,\-cm$^{-2}$.

The plateau width of the ${\nu}=1$ state has a {\it minimum} 
at the balanced point.
As the total electron density {\it increases},
the plateau width at the balanced point decreases. 
Eventually, the state at the balanced point disappears 
at the total density 
{\it more} than $1.5{\times}10^{11}$\,\-cm$^{-2}$.

The plateau width of the ${\nu}=2$ state has an intriguing behavior.
Around the balanced point,
its behavior is quite similar to that of the $\nu=2/3$ state.
Namely, it has a peak at the balanced point, 
and as the total density decreases,
the plateau width at the balanced point {\it decreases}. 
However, 
its behavior at off-balanced point ($|n_f-n_b|/n_t \gtrsim 0.2$)
is rather similar to that of the $\nu=1$ state.
The plateau width {\it increases} as the total density decreases.
Furthermore, when the total density is sufficiently small 
($n_t\simeq 0.6 \times 10^{11}$cm$^{-2}$),
the entire behavior now bears 
a close resemblance to that of the $\nu=1$ state.

The data of $\nu=2/3$ and 1 show clearly that 
the two QH states belong to two different types of QH states.
Moreover, 
the data of $\nu=2$ indicate that there are two types of $\nu=2$ QH state 
with different properties.

To confirm these observations we measured 
the activation energy $ \Delta $, which is derived 
from the temperature dependence of the magnetoresistance:
$R_{xx}=R_0 \exp (-\Delta /T)$.
As we will see, there is a close connection between the plateau width
and the activation energy.

In Fig.~\ref{UnbalancePS} we show the activation energy of 
the $\nu=1$ state and the $\nu=2$ state 
as a function of the density difference. 
As an example of the $\nu=1$ state we show the data 
when the total density is $1.1\times10^{11}$\,cm$^{-2}$. 
The activation energy is $\Delta=1$\,K at the balanced point
and gradually increases to $\Delta=2$\,K at $(n_f-n_b)/n_t$=-0.6. 
For the $\nu=2$ state we give the data 
for two values of the total density.
The activation energy at a lower density 
($0.6 \times 10^{11} $\,cm$^{-2}$)
is quite similar to that of the $ \nu =1 $ state, 
while the one at a higher density ($1.4\times10^{11} $\,cm$^{-2}$)
has an entirely different property: 
it has a peak at the balanced point.
The overall shapes of the activation energies are in good agreement
with the plateau widths seen in Fig.~\ref{Fig2PS}.

In Fig.~\ref{BalancePS} we show the activation energy of 
the $\nu=2/3$, 1 and 2 QH states
at the balanced point as a function of the total electron density.
The activation energy of the $\nu=2/3$ state shows a weak dependence
on the total electron density
and vanishes at $n_t$ less than $1.1 \times 10^{11}$\,cm$^{-2}$.
On the other hand, 
the activation energy of the $\nu=1$ state increases 
as the total electron density decreases 
and becomes almost constant 
for $n_t \leq 1.0 \times 10^{11}$\,cm$^{-2}$.
The activation energy of the $\nu=2$ state linearly 
depends on the total electron density 
larger than $0.9\times10^{11}$\,cm$^{-2}$,
and is almost constant when the total density is less.

Let us discuss the peculiar properties of the bilayer QH states
observed in our present data.
The basic nature of the QH state is governed by the competition 
between the intralayer and interlayer Coulomb interactions.
Both the Zeeman energy ($g^{*}\mu_B B$) 
and the tunneling energy ($\SAS$)
are much smaller than the Coulomb energy 
($e^2/\epsilon \ell_{B}$):
$g^{*}\mu_B B/(e^{2}/\epsilon\ell_B) \simeq 0.02$ and 
$\SAS/(e^{2}/\epsilon\ell_B) \simeq 0.04$ 
at $B=10$\,Tesla in our sample, 
where $\ell_B$ is the magnetic length.

We first consider the case where
the interlayer Coulomb interaction is negligible 
with respect to the intralayer Coulomb interaction.
In this case the bilayer system 
decouples into two degenerate independent monolayer systems. 
Consequently, the compound state becomes stable
at $\nu=1/m+1/m$ in the balanced configuration,
where all electron spins are polarized.
The tunneling interaction is suppressed 
and the excitation gap may involve mainly spin flips.
Even if the interlayer Coulomb interaction is not negligible,
the compound state will be realized 
since the monolayer $\nu=1/m$ state is very stable,
unless there exist a more stable state at this filling factor.

The $\nu=2/3$ state and a part of the $\nu=2$ state 
have clearly all the properties of the compound state.  
First, they are sharply enhanced in the balanced configuration 
as in Fig.~\ref{Fig2PS}.
Second, these state become unstable 
as the total electron density decreases
(or equivalently $d/\ell_B$ decreases).
Third, their activation energy in Fig.~\ref{BalancePS} behaves 
as in the monolayer case \cite{ActivMono},
though this is not so clear for the state at $\nu=1/3+1/3$.
More quantitatively, by a finite-size calculation \cite{SongHe},
the compound state at $\nu=1/3+1/3$ is known to be unstable 
when $d/\ell_B\leq1.5$.
In our experiment the peak of the $\nu=2/3$ state collapses
when $d/\ell_B\leq2.4$,
where $d$ is the interlayer separation.

Next, we discuss the case where the interlayer Coulomb interaction
is dominant.
In general, the bilayer QH state is described by 
the extended Laughlin wave function \cite{Laughlin}
$\Psi_{m_{\rm f}m_{\rm b}m}$ at
\begin{equation}
\nu  ={m_{\rm f}+m_{\rm b}-2m \over  
m_{\rm f} m_{\rm b}-m^2} \leq 1,
\end{equation}
where odd integers $m_{\rm f}$ and $m_{\rm b}$ represent 
the intralayer electron correlations,
while integer $m$ represents the interlayer correlation
induced by the interlayer Coulomb interaction.
(The compound state is obtained as a special limit $m=0$.)
The density ratio is fixed as
\begin{equation}
{n_{\rm f}\over n_{\rm b}}=
{m_{\rm b}-m \over m_{\rm f}-m} .
\label{FixedDensi}
\end{equation}
A strong interlayer correlation supports 
the ``coherent" state with $m_f=m_b=m$,
where the density ratio (\ref{FixedDensi}) becomes undetermined.
It is a characteristic feature of 
this state\cite{EIcoher,EIplasmon,SSC}
that it is stable at any density ratio
and that the IQC may develop spontaneously.

The $\nu=1$ state in our data can be identified 
as this ``coherent" state,
since the state continues to exist over all measured range of the
density difference, as in Fig.~\ref{Fig2PS}.
As the total density increases, 
or equivalently as $d/\ell_B$ increases,
the stability decreases as in Fig.~\ref{Fig2PS}, 
because the interlayer Coulomb interaction becomes weaker.
This can also be seen from the behavior of the activation energy 
in the balanced point in Fig.~\ref{BalancePS}.
The state may be regarded sufficiently stable 
for $n_t\leq1.0\times10^{11}$cm$^{-2}$,
where the activation energy is almost constant.
The QH state breaks down at and above the critical density of
$1.5\times10^{11}$\,cm$^{-2}$ ($d/\ell_B\simeq 2.2$),
which is in good agreement 
with the previous experiments\cite{DBLYexp}.

As found in Fig.~\ref{Fig2PS},
the ``coherent" state is least stable in the balanced configuration. 
This is also interpreted as an effect 
due to an interlayer Coulomb correlation.
The activation energy in Fig.~\ref{UnbalancePS} is understood 
if the excitation gap is dominated 
by the charging energy proportional to $(n_f-n_b)^2$.

The $\nu=2$ QH state undergoes a phase transition from 
a compound state to a ``coherent" state
as the total density decreases or the densities are unbalanced,
i.e.,
as the interlayer Coulomb interaction is increased over 
the intralayer Coulomb interaction.
The stability of the $\nu=2$ QH state 
in the vicinity of the balanced configuration 
is quite similar to that of the $\nu=2/3$ state, 
and it can be regarded as a compound state at $\nu=1+1$.
However, when the total density becomes sufficiently small 
at $n_t=0.6 \times 10 ^{11}$\,cm$^{-2}$,
we observe in Fig.~\ref{Fig2PS} and Fig.~\ref{UnbalancePS}
that the plateau width and the activation energy behave 
as those of the $\nu=1$ state
which we identify as the ``coherent" state.
The reason why we have a ``coherent" state at $\nu=2$ can be
explained by considering the spin degree of freedom.
In the $\nu=1$ state all electrons are in the {\it spin-up} states.
It is natural to expect another ``coherent" state 
with {\it spin-down} polarization 
above the $\nu=1$ state at $\nu=2$.
In this $\nu=2$ ``coherent" state 
the total system is spin unpolarized.
On the contrary,
the $\nu=1+1$ compound state is fully spin polarized.
This interpretation of the $ \nu =2 $ state is consistent with 
a recent inelastic light scattering experiment \cite{NuTwoExper}, 
where they observed a spin-polarized state in a high density sample 
and a spin-unpolarized state in a low density sample at $ \nu =2 $.

In conclusion, 
by measuring the plateau width and the activation energy
with continuously changing the electron density in each layer,
we have revealed that the observed bilayer QH states
($\nu=2/3$, 1 and 2) can be categorized 
into two distinctly different states.
The $\nu=2/3$ state is a typical compound state
with $\nu=1/3+1/3$ stabilarized solely by
the intralayer Coulomb interaction,
whereas the $\nu=1$ state is a ``coherent" state
stabilarized by the strong interlayer Coulomb interaction.
The $\nu=2$ state shows a phase transition from the compound state
to the ``coherent" state
as the interlayer Coulomb interaction becomes dominant.
The appearance of such $\nu=2$ ``coherent" state is a consequence
of the spin degree of freedom in the bilayer QH state.

We thank T. Saku (NTT) for growing the sample used in the present work,
T. Nakajima for useful discussions, 
and N. Kumada for experimental support.
Part of this work was done 
at Laboratory for Electronic Intelligent Systems,
RIEC, Tohoku University. 
Supports from a Grant-in-Aids for the Scientific Research 
from the Ministry of Education, Science, Sports and Culture
(08640438, 09244103, 09244204), 
and from the Multi-disciplinary Science Foundation are acknowledged.

\newpage
\begin{figure}[htb]
\caption{
Hall resistance versus magnetic field 
at various density difference 
at a fixed total electron density $n_t$.
The origins of the magnetic field axis are shifted 
in correspondence with the normalized density deference
$(n_{\rm f}-n_{\rm b})/n_{\rm t}$.}
\label{Fig1PS}
\end{figure}

\begin{figure}[htb]
\caption{
The Hall-plateau width of the ${\nu}=2/3$, 1 and 2 states 
at 50\,mK 
as a function of the electron density difference 
at several fixed total electron densities.
The lines are guides to the eye.}
\label{Fig2PS}
\end{figure}

\begin{figure}[t]
\caption{
Activation energy of the $\nu=1$ and 2 states 
as a function of the density difference.
The total density is fixed at a constant value. 
The curves with square data points are fitted by
assuming that the activation energy depends 
on $ (n_{\rm f}-n_{\rm b})^2 $.
}
\label{UnbalancePS}
\end{figure}

\begin{figure}[t]
\caption{
Activation energy of the $\nu=2/3,\ 1,\ 2$ QH states 
at the balanced density point, 
as a function of the total electron density.
The lines are guides to the eye.
}\label{BalancePS}
\end{figure}
\end{multicols}
\input epsf.tex
\epsfbox{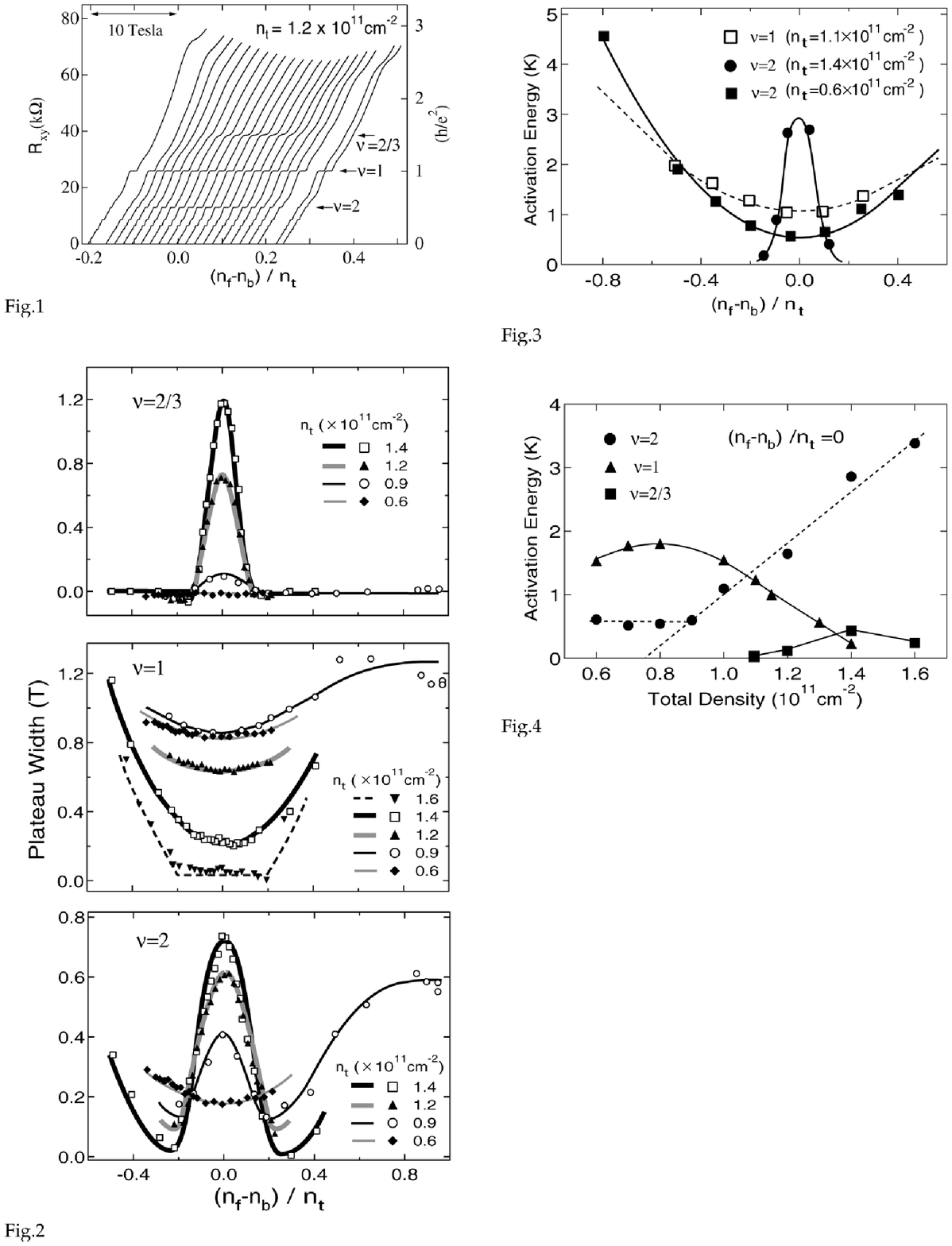}

\begin{references}
\bibitem{NoteA}
By the ``coherent" state we mean the Halperin $\Psi_{mmm}$ state,
upon which the IQC is expected to develop spontaneously.
\end{references}
\end{document}